\documentclass[twocolumn,amsmath,amssymb,superscriptaddress]{revtex4-2}

\usepackage{graphicx}

\usepackage{siunitx}
\usepackage{xspace}
\usepackage{hyperref}

\renewcommand{\bm}[1]{\boldsymbol{\mathbf{#1}}}
\newcommand{\ud}{\mathrm{d}}
\newcommand{\ie}{i.e.\@\xspace}
\newcommand{\bra}{\left\langle}
\newcommand{\ket}{\right\rangle}
\newcommand{\rg}{{\bf r}}
\newcommand{\nmode}{{n}_{ph}}
\newcommand{\DOS}{{\cal N}}

\begin{document}

% Title
% =====

\title{Photon thermalization in an open and disordered scattering medium}

\author{Lorenzo Soncin}
\affiliation{Institut Langevin, ESPCI Paris, Université PSL, CNRS, 1 rue Jussieu, 75005 Paris,
France}

\author{Romain Pierrat}
\affiliation{Institut Langevin, ESPCI Paris, Université PSL, CNRS, 1 rue Jussieu, 75005 Paris,
France}

\author{Yannick De Wilde}
\affiliation{Institut Langevin, ESPCI Paris, Université PSL, CNRS, 1 rue Jussieu, 75005 Paris,
France}

\author{Rémi Carminati}
\email{remi.carminati@espci.fr}
\affiliation{Institut Langevin, ESPCI Paris, Université PSL, CNRS, 1 rue Jussieu, 75005 Paris,
France}
\affiliation{Institut d’Optique Graduate School, Université Paris-Saclay, 91127 Palaiseau, France}

\author{Valentina Krachmalnicoff}
\email{valentina.krachmalnicoff@espci.fr}
\affiliation{Institut Langevin, ESPCI Paris, Université PSL, CNRS, 1 rue Jussieu, 75005 Paris,
France}

% Abstract
% ========

\begin{abstract}
   Thermalization of light, where photons acquire a temperature and chemical potential analogous to
   a material gas, remains a striking yet experimentally elusive manifestation of quantum
   statistical physics. To date, it has been realized only in carefully engineered photonic
   environments that enforce repeated absorption–emission cycles. Here we show that such
   thermalization can emerge in a radically simpler setting: an open scattering medium. Using a
   pumped fluorescent dye solution doped with colloidal particles, we demonstrate that multiple
   scattering alone suffices to trap photons long enough to drive them to thermal equilibrium. The
   emitted radiation follows a Bose–Einstein distribution with a finite chemical potential,
   independently tunable via optical pumping, while its temperature is set by the host medium. A
   clear spectroscopic signature of thermalization is observed as a plateau at the sample
   temperature over a finite spectral range. Our results establish disordered scattering media as a
   generic platform for photon thermalization, extending this fundamental phenomenon beyond resonant
   cavities and opening new routes towards cavity-free photonic thermodynamics and thermal light
   sources operating under ambient conditions.
\end{abstract}

%\keywords{Photon thermalization, Scattering, Random media, Fluorescence}

\maketitle

\section{Introduction}
% ====================

The electromagnetic field at thermodynamic equilibrium is characterized by a universal spectral
distribution known as Planck’s law~\cite{PLANCK-1914}. In a blackbody cavity, photons repeatedly
interact with matter through absorption and emission processes until they reach thermal equilibrium
with the cavity walls. From a statistical-physics perspective, the resulting radiation can be viewed
as a gas of bosons obeying Bose–Einstein statistics with zero chemical potential, reflecting the
fact that the photon number is not conserved and is determined solely by temperature. In this
context, producing thermal visible light without heating the system to temperatures of several
thousands degrees is particularly challenging. 

A fundamentally different situation arises when photons repeatedly interact with externally pumped
emitters: the resulting radiation can reach thermal equilibrium while maintaining a finite chemical
potential~\cite{WURFEL-1982}. In this regime, the radiation field is described by a Bose–Einstein
distribution whose temperature is set by the environment, whereas the chemical potential can be
controlled independently through optical or electrical pumping. Such thermalized photon gases have
attracted considerable interest because they provide a route toward phenomena such as photon
Bose–Einstein condensation and controlled photonic thermodynamics.

\begin{figure*}[t]
   \centering
   \includegraphics[width=0.75\linewidth]{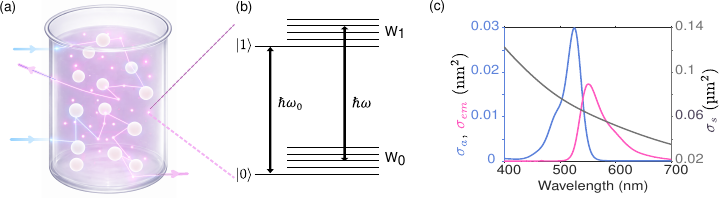}
   \caption{Active disordered scattering medium for photon thermalization. (a)
   Schematic view of the active scattering medium. Colloidal polystyrene beads are dispersed in an
   aqueous solution containing Rhodamine 6G fluorescent dye molecules, creating an open disordered
   medium that confines emitted photons through multiple scattering. (b) Jablonski diagram
   of the molecular levels. (c) Absorption and emission cross sections of the active medium
   (blue and pink respectively) and scattering cross section of the disordered medium.}
   \label{fig1}
\end{figure*}

Despite their fundamental importance, experimental realizations of photon thermalization remain
remarkably limited. To date, they have relied on carefully engineered photonic environments designed
to enforce repeated absorption–emission cycles, including optical cavities~\cite{KLAERS-2010-1,
MARELIC-2016, DUNG-2017, BUSLEY-2023}, erbium-doped fibers~\cite{WEILL-2017, FERRARO-2024},
electrically pumped semiconductors microresonators and quantum wells~\cite{BARLAND-2021,
SCHOFIELD-2024, PIECZARKA-2024}, or surface plasmons on lattices~\cite{HAKALA-2018}. This raises the
fundamental question of whether thermalization can emerge in an open and disordered environment,
without relying on the well-defined modes provided by resonant cavities and other highly engineered
photonic architectures.

Multiple scattering in disordered media provides a natural candidate. When the system size exceeds
the scattering mean free path, photons perform a diffusive random walk that substantially increases
their residence time within the medium, effectively trapping light despite the open nature of the
system. This mechanism is well known to enhance light–matter interactions and constitutes the basis
of random lasing in systems ranging from colloidal dye suspensions to infiltrated powders and white
paints~\cite{CAO-2005, WIERSMA-2008}.

Below the random-lasing threshold, however, the role of multiple scattering has remained largely
unexplored. In this regime, scattering may provide sufficient photon trapping to promote repeated
absorption and emission events while avoiding the onset of coherent amplification. One may therefore
expect a transition between the weak-scattering regime, where scattering mainly enhances light
extraction without modifying emitter dynamics~\cite{SHIN-2015,VASKIN-2019,YALCIN-2020}, and the
random-laser regime, where gain dominates. Here we show that this intermediate regime hosts a
fundamentally different phenomenon: the thermalization of light in an open disordered medium. We
demonstrate that multiple scattering alone can drive the emitted radiation toward a Bose–Einstein
distribution with a finite, pump-controlled chemical potential, establishing disordered scattering
media as a generic platform for photon thermalization beyond conventional architectures which must
be carefully designed to select only a few modes. We show a substantial modification in the spectrum
of the radiation emitted by a systems mixing Rhodamine 6G molecules with scattering polystyrene
beads under low pump excitation, \ie equivalent to sunlight.

By increasing the beads concentration, the emission spectrum exhibits a pronounced red shift due to
increased scattering, and adopts a Bose–Einstein distribution with a finite chemical
potential~\cite{HERRMANN-2005}.  Using an original data-analysis approach, we demonstrate that the
photon gas thermalizes over a spectral window exceeding \SI{20}{nm}, coinciding with the overlap
between the emission and absorption cross-sections of the dye. Thermalization occurs at the sample
temperature, independent of the chemical potential, which is instead controlled by the pump
intensity. A direct comparison with a fluorescent Rhodamine 6G sample in the absence of scatterers
strikingly highlights the essential role of multiple scattering in enabling thermalization.

\section{Thermalization process}
% ==============================

The system under study consists of an active medium of dye molecules in water and a scattering
medium made of colloidal polystyrene beads (PolyScience microspheres, nominal diameter
\SI{370}{nm}), pumped by a laser beam as shown in Fig.~\ref{fig1}\,(a). The fundamental and excited
electronic states $|0\rangle$ and $|1\rangle$ of the molecules, together with their rovibrational
sublevels, are schematically represented in Fig.~\ref{fig1}\,(b). Since the lifetime of the excited
electronic state (a few ns) is large compared to the collision time of the rovibrational states
($<1$ ps), the latter are assumed to be at thermal equilibrium at the temperature $T$ of the
solution (room temperature). In this situation the absorption and stimulated emission cross sections
$\sigma_a$ and $\sigma_{em}$ satisfy the Kennard-Stepanov relation~\cite{KENNARD-1918,STEPANOV-1957}
\begin{equation}
   \frac{\sigma_{em}(\omega)}{\sigma_a(\omega)} = \frac{W_0}{W_1} \exp\displaystyle\left[\frac{-\hbar(\omega-\omega_0)}{k_BT}\right ] \, ,
   \label{eq:Kennard-Stepanov}
\end{equation}
where $\omega_0$ is the Bohr frequency of the electronic transition, $k_B$ is the Boltzmann constant
and $\hbar$ the reduced Planck constant. In this expression $W_i =\int_0^\infty g_i(E) \exp(-E/k_BT)
\ud E$ is the partition function of the rovibrational levels in the electronic state $|i\rangle$, with
$g_i(E)$ their density of states, and acts as a weighting factor for state $|i\rangle$. The
absorption and emission cross sections of the active medium and the scattering cross section of the
polystyrene beads are displayed in Fig.~\ref{fig1}\,(c).  

\begin{figure*}[t]
   \centering
   \includegraphics[width=0.75\linewidth]{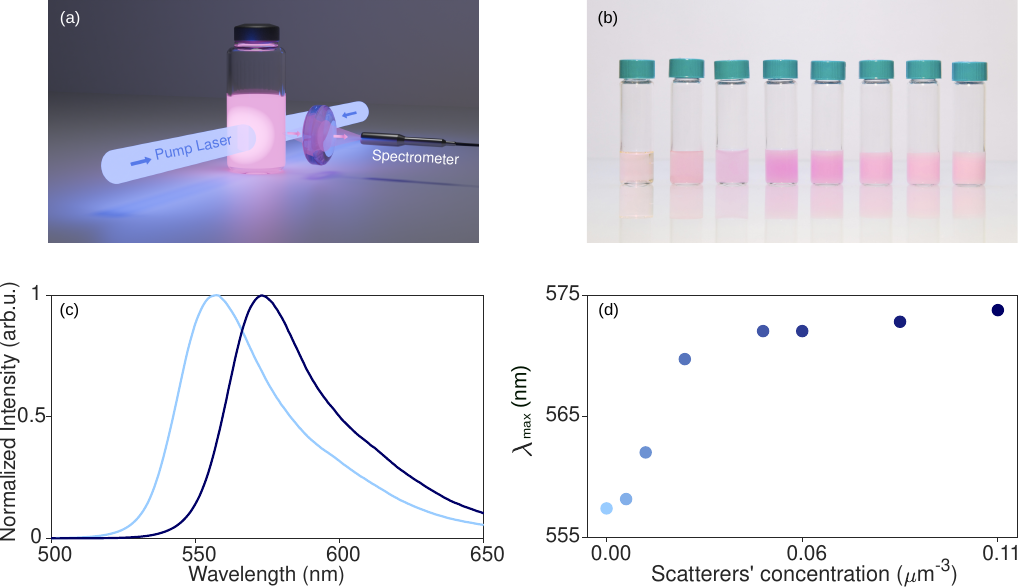}
   \caption{Influence of scatterers concentration on the emission spectrum. (a)
   Artist view of the experimental setup. (b) Image of samples with a fixed concentration of
   Rhodamine 6G and variable concentration of scatterers. (c) Emission spectra of aqueous
   Rhodamine 6G solutions with and without scatterers (respectively dark blue and light blue solid
   lines). The Rhodamine 6G concentration was fixed at \SI{5.36e3}{\micro m^{-3}} for both spectra,
   while the scatterer concentration varies from \num{0} to \SI{0.11}{\micro m^{-3}}. (d)
   Position of the maximum of the emission spectrum as a function of the scatterer concentration.
   The spectrum is red shifted when the concentration of scatterers increases and saturates for a
   density of scatterers of \SI{0.11}{\micro m^{-3}}.}
   \label{fig2}
\end{figure*}

In our experiments, the colloidal beads dispersed in the dye solution create a disordered medium
that confines the emitted photons by multiple scattering. Modelling {\it ab initio} the interplay
between wave scattering and the quantum dynamics of the emitters is an ambitious challenge. In the
simplest picture, the system can be described by coupling rate equations for the populations of the
electronic states with a transport equation governing the average photon density in the scattering
medium. When the system size $R$ largely exceeds the photon transport mean free path $\ell_t$, the
average photon density obeys a diffusion equation with diffusion constant $D=v_E \ell_t/3$, where
$v_E$ is the energy velocity (for non-resonant scattering $v_E \simeq c/n_r$, where $n_r$ is the
real part of the refractive index of the suspension and $c$ is the speed of light in
vacuum)~\cite{Akkermans_Montambaux_2007,CARMINATI-2021-1}. The diffusion model is relevant to
introduce thermalization by multiple scattering in simple terms, and to support the analysis of the
experiments presented in this work, but a more refined transport model can be used if needed (see
App.~\ref{app_theory}). Denoting by $n(\rg,\omega)$ the photon density, such that $n(\rg,\omega)
\ud\omega$ is the number of photons per unit volume at point $\rg$ with frequency in
$[\omega,\omega+\ud\omega]$, we can write
\begin{multline}
   \frac{\partial \, n(\rg,\omega)}{\partial t} - D \nabla^2 n(\rg,\omega) = - \rho_0 \, \sigma_a(\omega) v_E \, n(\rg,\omega) 
\\
   + \rho_1 \, \sigma_e(\omega) v_E \, n(\rg,\omega)+  \rho_1\sigma_{em}(\omega) v_E \,  \DOS(\omega)  \, ,
  \label{eq:diffusion}
\end{multline}
where $\rho_0$ and $\rho_1$ are the populations of the electronic states $|0\rangle$ and
$|1\rangle$, respectively, and $\DOS(\omega)$ is the photon density of states in the medium. The
three terms in the right-hand side represent absorption by the molecules, stimulated emission, and
spontaneous emission, respectively.  Similar models coupling multiple scattering of light with a
pumped medium have been used in the context of random
lasers~\cite{WIERSMA-1995,PIERRAT-2007,FROUFE-2009}. Here we assume a spatially uniform pumping, \ie
we neglect the influence of scattering on the excitation laser. Moreover, since $\rho_0$ and $\rho_1$ are weakly affected
by absorption and emission of the (non-pumping) photons, they are considered to be uniform in the
model.

Equation~\eqref{eq:diffusion} suggests a simple qualitative analysis. We introduce the effective
absorption time $\tau_a^\star = [\rho_0 \, \sigma_a v_E - \rho_1 \sigma_{em}v_E]^{-1}$ as the
absorption time corrected for stimulated emission, and the diffusion time $\tau_d = \alpha R^2/D$
across the scattering medium, where $\alpha$ is a numerical prefactor that depends on the geometry
($\alpha \simeq 1/6$ for a cylindrical cuvette with radius $R$). Thermalization is expected in the
regime $\tau^\star_a \ll \tau_d$ in which photons can be absorbed and emitted by the dye molecules
over many cycles during the trapping time by multiple scattering. In this regime, we can neglect the
diffusion term in the left-hand side in Eq.~\eqref{eq:diffusion}. In steady-state, this immediately
leads to a photon density $n(\rg,\omega) = \nmode(\omega) \DOS(\omega)$, with the average density of
photons per mode $\nmode(\omega)$ satisfying 
\begin{equation}
   \nmode(\omega) = \frac{1}{\displaystyle\frac{\rho_0 \sigma_a(\omega)}{\rho_1 \sigma_{em}(\omega)}-1} \, .
   \label{eq:nph_steady_interm}
\end{equation}
The average ratio of excited molecules $\rho_1/\rho_0$ being fixed by external pumping, it is useful
to introduce the photon chemical potential 
\begin{equation}
   \mu = \hbar\omega_0 + k_B T \ln \frac{\rho_1 W_0}{\rho_0 W_1}  \, .
   \label{eq:chemical_potential}
\end{equation}
We note that $\mu$ is similar to the chemical potential introduced in Ref.~\cite{WURFEL-1982} in the
context of light emission by semiconductors. Making use of Eqs.~\eqref{eq:Kennard-Stepanov},
\eqref{eq:nph_steady_interm} and \eqref{eq:chemical_potential}, we find that the average density of
photons per mode obeys Bose-Einstein statistics
\begin{equation}
   \nmode(\omega) = \frac{1}{\exp\displaystyle\left(\frac{\hbar\omega-\mu}{k_BT}\right )-1} \, ,
   \label{eq:nph_steady}
\end{equation}
showing that, under the condition $\tau^\star_a \ll \tau_d$, the photon gas inside the medium is
thermalized with a non-zero chemical potential. In the experiment, we expect to observe the
signature of thermalization in the spectrum of the observable signal $I(\omega)$ formed by diffuse
light leaking through the medium boundaries, that takes the form
\begin{equation}
   I(\omega) = \frac{\eta \, \hbar\omega  \,\DOS(\omega)}{\exp\displaystyle\left(\frac{\hbar\omega-\mu}{k_BT}\right )-1} \, ,
   \label{eq:exp_intensity}
\end{equation}
with $\eta$ a prefactor that depends on the geometry and the efficiency of the photodetection chain.
It is interesting to note that in steady-state the condition for thermalization $\tau^\star_a \ll
\tau_d$ can be rewritten in terms of length scales. Introducing the effective absorption length
$\ell_a ^\star = v_E \tau_a ^\star$, thermalization is reached when $\ell_a^\star \ll 3 \alpha
R^2/\ell_t$. An approach to define the thermalization regime in terms of length scales is provided
in App.~\ref{app_theory}.

\section{Experimental evidence of thermalization}
% ===============================================

The active sample, consisting of Rhodamine 6G molecules and polystyrene beads in water, is
illuminated with two counter-propagating collimated laser beams at a wavelength of \SI{473}{nm}, as
shown in Fig.~\ref{fig2}\,(a). The two beams have equal power and a spatial extent comparable with the
diameter of the sample's flask (\SI{15}{mm}), ensuring uniform pumping across the sample. The beam
polarizations are crossed to prevent interferences in the excitation field. The emitted radiation is
collected through an optical fiber and dispersed in a spectrometer (see App.~\ref{app_setup} for
details on the experimental setup). Optics between the sample and the fiber enable the collection of
the radiation emitted over a numerical aperture NA= \num{0.22} from a region with size
\SI{50}{\micro m}, and symmetrically positioned with respect to the excitation beams. The fiber
input is conjugated with the outer surface of the glass flask. 

Figure~\ref{fig2}\,(b) is a picture of the different samples under study, with increasing levels of
scattering controlled by varying the density of polystyrene beads (see App.~\ref{app_sample} for
details on sample preparation). From left to right the density of scatterers increases from zero to
\SI{0.11}{\micro m^{-3}} at a constant density of Rhodamine 6G fixed at \SI{5.36e3}{\micro m^{-3}}.
The measured spectral properties are reported in Fig.~\ref{fig2}\,(c) and \ref{fig2}\,(d).  When the
density of scatterers is increased from zero to \SI{0.11}{\micro m^{-3}}, the spectrum of the
emitted radiation is red-shifted, as shown in Fig.~\ref{fig2}\,(c). This can be easily understood as
an effect of progressive reabsorption of the radiation emitted by the molecules in the spectral
region in which absorption and emission cross sections overlap. Interestingly, Fig.~\ref{fig2}\,(d)
shows that the intensity maximum is not only red shifted but saturates to a value of \SI{575}{nm}
when the density of scatterers increases.  Saturation of the spectral shift is a crucial ingredient
for photon thermalization as it indicates the occurence of emission and absorption by the molecules
within the residence time of the photons in the medium.

\begin{figure}[t]
   \centering
   \includegraphics[width=\linewidth]{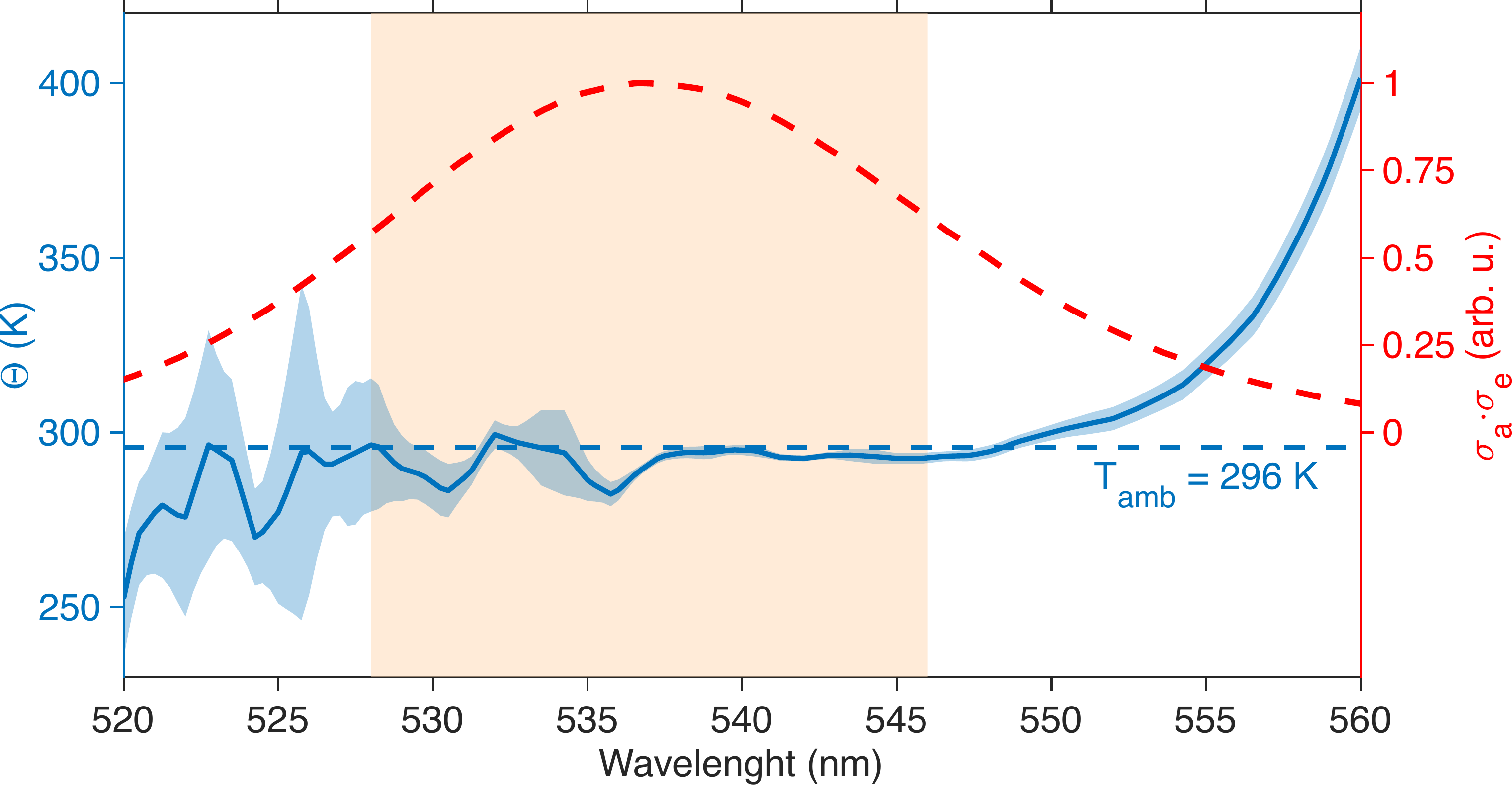}
   \caption{Experimental signature of thermalization. Blue solid line: Value of
   $\Theta$ extracted from the thermalized emission spectra of an aqueous solution containing
   Rhodamine 6G (\SI{5.36e3}{\micro m^{-3}}) and scatterers (\SI{0.11}{\micro m^{-3}}). The blue
   shadowed area indicates the error bars. The orange-shaded area highlights the region where the
   plateau is expected to appear. Red dashed line: Product of the absorption and emission cross
   sections of the Rhodamine 6G.}
   \label{theta_T_amb}
\end{figure}

\subsection*{Criterion for thermalization}
% ========================================

To define a criterion for thermalization that is robust in practice, we introduce $S(\omega)
=I(\omega)/(\hbar \omega \mathcal{N}(\omega))$, where $I(\omega)$ is the collected intensity and
$\mathcal{N}(\omega)=\omega^2 /(\pi^2 v_E^3)$ is the average photon density of states in the medium.
For thermalized photons at a temperature $T$, $I(\omega)$ follows Eq.~\eqref{eq:exp_intensity}.  In
the current experiment, at $T = 300\,\mathrm{K}$, in correspondence of the maximum overlap between
the absorption and emission spectrum of Rhodamine 6G, we have $\hbar\omega \simeq 2.3\,\mathrm{eV}$
and $k_B T \simeq 0.025\,\mathrm{eV}$. Since $\mu < 0$, then $\hbar\omega - \mu \gg k_B T$, and
therefore we find that $\ln S(\omega) \simeq  \ln \eta- (\hbar\omega-\mu)/(k_B T)$.  Assuming that
the collection coefficient $\eta$ is independent of $\omega$, this suggests to define the parameter
\begin{equation}
   \Theta(\omega) = - \frac{\hbar}{k_B} \left[\left. \frac{\mathrm{d}\ln S(\omega)}{\mathrm{d}\omega} \right. \right]^{-1}
   \label{def:theta}
\end{equation}
that equals the temperature $T$ of the sample when photon thermalization is achieved. In practice,
in the thermalized regime, the plot of $\Theta (\omega)$ is expected to exhibit a plateau at a value
$\Theta (\omega)=T$ in the spectral region where the absorption and emission cross-sections of the
molecules overlap.

\begin{figure*}[t]
   \centering
   \includegraphics[width=\linewidth]{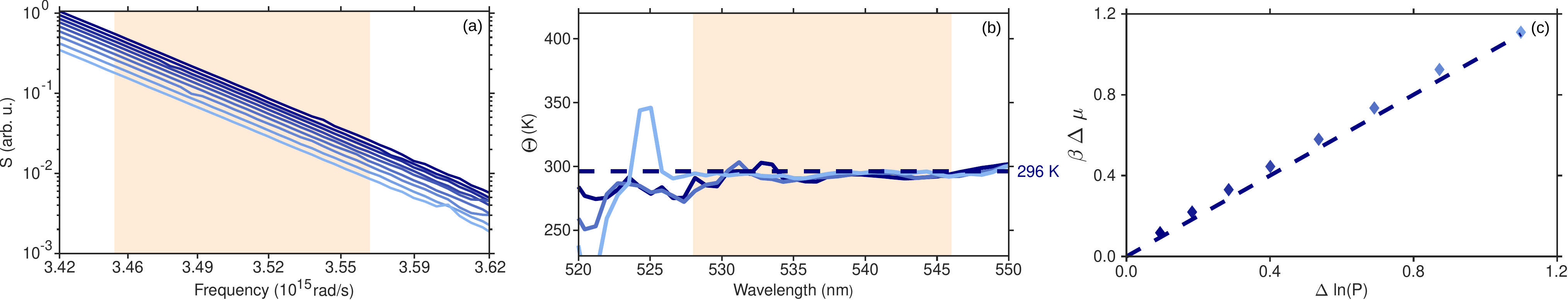}
   \caption{Experiments with variable power. (a) Experimental measurement of
   $S(\omega)$ for pump powers going from \SI{8}{mW} to \SI{24}{mW} in steps of \SI{2}{mW}. Darker
   shades of blue correspond to higher powers. The spectra are normalized  with respect to the
   maximal intensity of the one obtained with the highest pump power. The orange shaded region
   highlights as in Fig.~\ref{theta_T_amb} the range where thermalization is expected. (b)
   Plot of $\Theta(\lambda)$ for three representative spectra shown in panel (a), respectively for
   \SI{8}{mW} (light blue), \SI{14}{mW} (mild blue), and \SI{24}{mW} (dark blue). The dashed blue
   line represents ambient temperature. (c) Variation of the chemical potential as a function
   of the variation of the logarithm of the pump power (blue dots). The colours of the dots
   correspond to the data shown in panel (a) and (b). The reference power is chosen at a value
   $P_{\text{ref}}=\SI{24}{mW}$. The dashed blue line is the theoretical expectation.}
   \label{theta_P_var}
\end{figure*}

The parameter $\Theta$ is plotted in Fig.~\ref{theta_T_amb} versus the wavelength $\lambda=2\pi
c/\omega$ (blue solid line) for a sample in which scatterers (concentration \SI{0.11}{\micro
m^{-3}}) are added to an aqueous solution of Rhodamine 6G (concentration \SI{5.36e3}{\micro
m^{-3}}). With these concentrations, for $\lambda =\SI{537}{nm}$, at which the overlap between the
emission and the absorption spectra is maximum, the absorption time is $\tau_a \simeq \tau_a^\star =
\SI{36}{ps}$ and the diffusion time is $\tau_d = \SI{213}{ps}$. The thermalization condition $\tau_a
\ll \tau_d$ is therefore satisfied. At this wavelength, the scattering and the transport mean free
path are $\ell_s=\SI{136}{\micro m}$ and $\ell_t=\SI{587}{\micro m}$ respectively, and are much
smaller than the size of the cuvette.  A plateau at the sample's temperature $T=\SI{296}{K}$ is
clearly visible in the spectral region ranging from $\lambda\simeq\SI{528}{nm}$ to $\lambda \simeq
\SI{546}{nm}$ indicated by the orange-shaded region in the figures. The appearance of the plateau is
the first direct evidence of photon thermalization in a scattering active sample. The blue shadowed
area represents the error bars calculated as the standard deviation from the average of \num{10}
measurements on the same sample (see App.~\ref{app_data} for the data analysis). Since between two
consecutive measurements the sample was removed and replaced on the sample holder, error bars
attest the good reproducibility of the measurement. At short wavelengths, the signal is noisy with
larger error bars due to the low emission intensity at the tail of the spectrum. Moreover, applying
the numerical treatment to extract $\Theta$ in a low signal-to-noise region further amplifies the
noise. To further stress that the presence of a plateau for $\Theta$ is a clear signature of
thermalization, we report in Fig.~\ref{theta_T_amb} the product of the absorption and emission
cross-sections measured in a neat solution of Rhodamine 6G without scatterers (red dashed line). The
maximum of the product of the cross sections coincides with the spectral region of the plateau.
According to this observation, experimental data deviate from room temperature when the product of
the cross sections falls below a critical value disrupting the balance between emission and
absorption events.  

\subsection*{Variation of the chemical potential}
% ===============================================

The chemical potential defined in Eq.~(\ref{eq:chemical_potential}) depends only on the ratio
$\rho_1/\rho_0$ of the populations of the excited and ground states of the molecules. Since $\rho_1$
is mostly driven by the the pump intensity, the chemical potential is considered to be an external
parameter driven by the pumping level. As a consequence, the dependence of the emision spectra on
the chemical potential can be checked by varying the laser power. The experimental results are shown
in Fig.~\ref{theta_P_var} for both the normalized emitted spectrum $S(\omega)$ [plotted in
logarithmic scale in panel (a)] and the parameter $\Theta(\lambda)$ [panel (b)], and for different
pump intensities ranging from \SI{8}{mW} to \SI{24}{mW}. The sample was held at ambient temperature
for all measurements. Since increasing the pumping intensity amounts to increasing $\rho_1/\rho_0$
and therefore the chemical potential $\mu$, we expect a vertical shift in $\ln S(\omega)$ for
increasing pumping power in the thermalized regime. This is clearly seen in the experimental data in
Fig.~\ref{theta_P_var}\,(a).  Moreover, the plot of $\Theta(\lambda)$ in Fig.~\ref{theta_P_var}\,(b)
reveals a key feature of the thermalization mechanism: For each pumping level a stable plateau at
ambient temperature $T$ is reached, demonstrating that $T$ and $\mu$ act as independent parameters
in the control of the emission spectrum. These results also confirm that the condition of uniform
chemical potential, which is facilitated by the chosen pumping geometry, appears to be met. 

A further hallmark of the thermalized regime is that, by measuring $\ln S(\omega)$ for two different
pumping powers, namely $P$ and a reference power $P_{\text{ref}}$, and deducing the chemical
potentials $\mu$ and $\mu_{\text{ref}}$, a linear relation $\ln P-\ln{P_{\text{ref}}} = \beta
(\mu-\mu_{\text{ref}})$ holds for a thermalized photon gas. The result obtained from the same data
as in Fig.~\ref{theta_P_var}\,(a,b) is plotted in Fig.~\ref{theta_P_var}\,(c), together with the
straight dashed line corresponding to the linear behavior expected from the theoretical relation.
The excellent agreement between the experimental data and the theoretical expectation, obtained
without fitting parameter, is another clear signature of thermalization.

\subsection*{Dependence on temperature}
% =====================================
\begin{figure*}[t]
   \centering
   \includegraphics[width=0.75\linewidth]{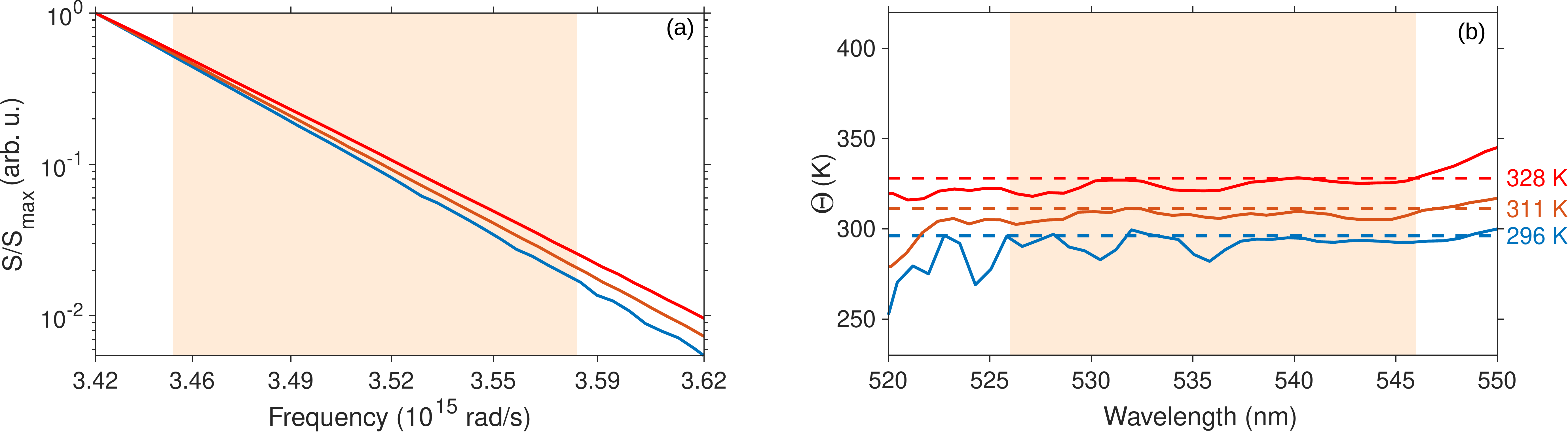}
   \caption{Experiments with variable temperature. (a) Experimental measurement
   of $S(\omega)$ normalized with respect to its maximum value, as the sample temperature was
   increased (blue, orange and red solid lines correspond to a temperature of \SI{296}{K},
   \SI{311}{K}, and \SI{328}{K} respectively). The orange-shaded region highlights the spectral
   range where thermalization occurs. (b) Plot of $\Theta (\lambda)$ for the spectra shown
   in panel (a). The temperature at which the sample was held during each measurement is indicated by
   the dashed lines.}
   \label{theta_T_var}
\end{figure*}

The second parameter that can be tuned experimentally is the temperature of the sample. To this aim,
a flat thermoresistance is installed below the sample, and the sample is heated until a steady-state
at the targeted temperature is reached. Measurements are reported in Fig.~\ref{theta_T_var}\,(a)
for $S(\omega)$ and in Fig.~\ref{theta_T_var}\,(b) for $\Theta(\lambda)$ for three different
temperatures of the sample, namely \SI{296}{K} (blue), \SI{311}{K} (orange), and \SI{328}{K} (red).
From the analytical expression of $S(\omega)$, an increase in temperature is expected to reduce the
slope of the spectrum, which is clearly observed in the experimental date plotted in
Fig.~\ref{theta_T_var}\,(a).  Moreover, when the parameter $\Theta(\lambda)$ is plotted for the same
sets of data, we observe a thermalization plateau at the sample temperature measured directly on the
sample with a thermocouple. As in Fig.~\ref{theta_P_var}, the error bars are not shown for clarity,
but they are consistent with the error bars plotted in Fig.~\ref{theta_T_amb} and commented
previously.

\begin{figure*}[t]
   \centering
   \includegraphics[width=0.75\linewidth]{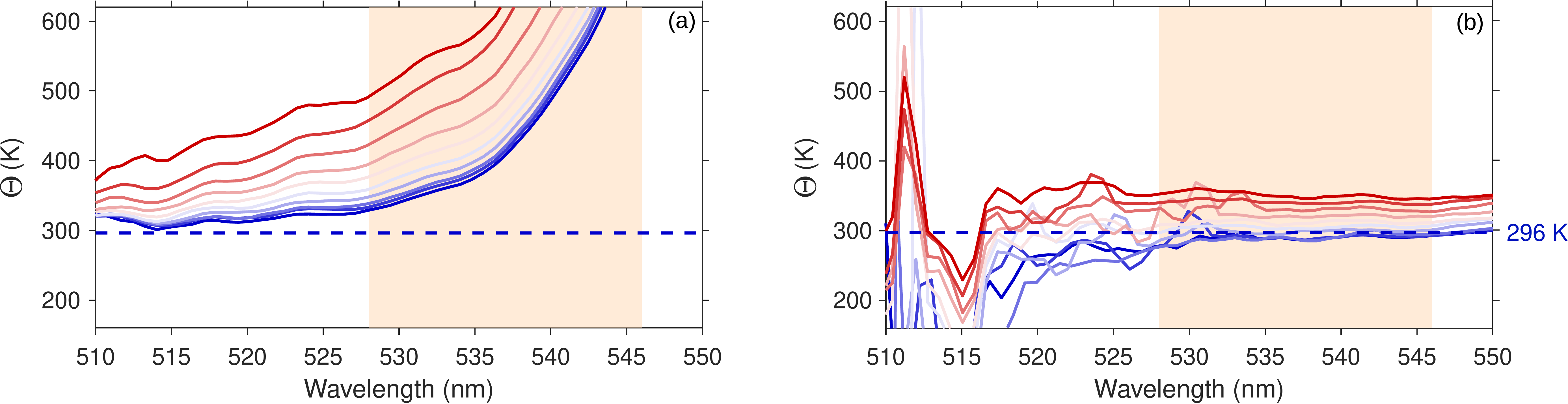}   
   \caption{Comparison between fluorescence and thermalization. (a) Plot of
   $\Theta(\lambda)$ extracted from the emission spectra of an aqueous solution of Rhodamine 6G as
   its temperature is increased from ambient temperature up to \SI{356}{K} (blue to red solid
   lines). The dashed blue line indicates the ambient temperature. (b) Plot of
   $\Theta(\lambda)$ extracted from the emission spectra of an aqueous solution of Rhodamine 6G with
   scatterers, measured as the sample temperature was increased from ambient temperature up to
   \SI{356}{K} [blue to red solid lines, as in panel (a)]. The orange shadowed region highlights the
   spectral range were thermalization is expected.}
   \label{theta_Q_notQ}
\end{figure*}

The role of multiple scattering in enabling photon thermalization becomes remarkably clear when we
compare the modification of the spectrum of the emitted light with temperature for samples without
scatterers (fluorescence regime) and with scatterers (thermalized regime). The results in terms of
the parameter $\Theta(\lambda)$ are shown in Fig.~\ref{theta_Q_notQ} for the sample without
scatterers [panel (a)] and with scatterers [panel (b)], for temperatures increasing from ambient
temperature to \SI{356}{K} (dark blue to red solid lines). As expected, since without scatterers the
photon gas is not thermalized, the parameter $\Theta(\lambda)$ in Fig.~\ref{theta_Q_notQ}\,(a) has a
shape that is qualitatively different from the data obtained in the presence of scattering, and
never reaches a plateau in the orange-shaded region. At ambient temperature (dark blue line) the
lowest value of $\Theta$ that is reached exceeds room temperature by about \SI{50}{K} (blue dashed
line). The spectral shape is also strongly modified when the temperature increases. On the contrary,
in the presence of sufficiently strong scattering to reach thermalization
[Fig.~\ref{theta_Q_notQ}\,(b)], $\Theta(\lambda)$ exhibits a plateau at a level that increases with
temperature (dark blue to red solid lines), in agreement with the previous observation that the
thermalization plateau coincides with the temperature of the sample.

\section{Conclusion}
% ==================

In conclusion, we have provided the first demonstration of photon thermalization with non-zero
chemical potential by multiple scattering in a remarkably simple open system, made of an aqueous
solution of pumped dye molecules mixed with a precisely tuned density of scatterers. Such a 
system, that is easily scalable and extremely low cost, reveals a new paradigm for photon
thermalization and Bose-Einstein condensation. Beyond establishing a new experimental and
theoretical platform, our results identify multiple scattering as a generic physical mechanism
capable of mediating thermal equilibrium between light and matter in open systems. This considerably
broadens the class of photonic environments in which equilibrium statistical mechanics can be
explored, extending concepts previously associated with cavities and other highly engineered
architectures to disordered media.

Our results also open new avenues for the development of thermal light sources capable of producing
large photon fluxes in the visible while operating close to ambient temperature. Fluorescent
scattering media offer a versatile platform for creating visually vivid, color-tunable surfaces
with tailored thermal properties and minimal absorption, providing an alternative to conventional
pigments and plasmonic nanoparticles. The multiple degrees of freedom afforded by such active
complex media enable independent control over absorption, reflection and transmission of the
incident solar spectrum at pump irradiances comparable to solar intensity, making them particularly
attractive for applications in photovoltaics and radiative cooling, where optical and thermal
functionalities must be engineered independently.

\section*{Data availability}
%===========================

All data that support the plots within this paper and other findings of this study are available
from the corresponding authors upon request.

\section*{Acknowledgements}
% =========================

This work was supported by the ``Investissements d’Avenir'' program launched by the French
Government (Labex WiFi) and by the Agence Nationale de la Recherche (FLUOSCATT Project No.
ANR-25-CE57-0642-01). LS was supported by a PhD fellowship from École Doctorale
Physique-en-Île-de-France (EDPIF).

The authors are grateful to J. Don Jayamanne, K. Chevrier and M. Vernet for their implication and
the fruitful discussions in the early stages of the project.

\section*{Author contributions}
% =============================

RC proposed the initial idea ; VK, YDW, RP, RC designed the experiments. LS, VK, YDW carried out the
experiments. LS and RP performed numerical simulations guiding the experiments and supporting the
analysis. All authors analysed the experimental data and discussed the results. All authors wrote
the paper.

\appendix

\section{Alternative theoretical model}\label{app_theory}
% =====================================

In the main text, we have considered a simple model based on a diffusion equation to derive criteria
for photon thermalization in strongly-scattering disordered media. Here we present an alternative
and more accurate model based on a generalized Radiative Transfer Equation (RTE). The RTE is useful
to describe the transport of the average intensity beyond the validity of the diffusion
approximation~\cite{CHANDRASEKHAR-1950,CARMINATI-2021-1}. The generalization consists in accounting
for the coupling with dye molecules, by adding terms corresponding to absorption, spontaneous
emission and stimulated emission. In the steady state regime, the full equation reads
\begin{multline}\label{rte}
   \bm{u}\cdot\bm{\nabla}_{\bm{r}}I(\bm{r},\bm{u},\omega)
      =-\rho_0\sigma_a(\omega)I(\bm{r},\bm{u},\omega)
       +\rho_1\sigma_{em}(\omega)I(\bm{r},\bm{u},\omega)
\\
       +\rho_0\sigma_{em}(\omega)v_E\frac{\mathcal{N}(\omega)}{4\pi}
       -\frac{1}{\ell_s(\omega)}I(\bm{r},\bm{u},\omega)
\\
       +\frac{1}{\ell_s(\omega)}\int_{4\pi}p(\bm{u}\cdot\bm{u}',\omega)I(\bm{r},\bm{u}',\omega)\ud\bm{u}'.
\end{multline}
Here $I(\bm{r},\bm{u},\omega)$ is the specific intensity which is a local (position $\bm{r}$) and
directional (unit vector $\bm{u}$) radiative flux at frequency $\omega$, and $\sigma_a$,
$\sigma_{em}$ and $\mathcal{N}(\omega)$ are, respectively, the absorption and emission cross
sections of the molecules and the average density of optical states. $\rho_0$ and $\rho_1$ are the
volume densities of molecules in the fundamental and excited states. The last two terms in
Eq.~\eqref{rte} describe the scattering processes, with $\ell_s$ the scattering mean free path and
$p(\bm{u}\cdot\bm{u}',\omega)$ the phase function describing the angular distribution of the
scattered intensity~\cite{CHANDRASEKHAR-1950,CARMINATI-2021-1}. A feature of the RTE is its ability
to describe the full range of scattering regimes, from ballistic to diffusive transport.

It should be noted that in Eq.~\eqref{rte} we have neglected intrinsic absorption in the scatterers.
Furthermore, the pumping of the molecules by the external laser is considered to be uniform such
that $\rho_1$ does not depend on position. We also stress that the specific intensity is normalized
by $\hbar\omega$ such that $v_E^{-1}\int_{4\pi} I(\bm{r},\bm{u},\omega)\ud\bm{u}$ has to be
understood as the number density of photons inside the scattering medium, with $v_E$ the energy
velocity.

In order to highlight the thermalization regime and the conditions for its existence, it will prove
useful to introduce the integral form of the RTE. This form of the RTE is at the root of the
celebrated Monte Carlo numerical scheme~\cite{MODEST-1993}. To this end, we introduce the effective
extinction mean free path
\begin{equation}
   \frac{1}{\ell_e^\star(\omega)}=\frac{1}{\ell_s(\omega)}+\rho_0\sigma_a(\omega)-\rho_1\sigma_{em}(\omega)
\end{equation}
and the source term $S(\omega)=\rho_1\sigma_{em}(\omega)v_E\mathcal{N}(\omega)/(4\pi)$. Considering
an infinite medium, we can easily take the Fourier transform of Eq.~\eqref{rte} with respect to the
space variable $\bm{r}$, which leads to
\begin{multline}
   I(\bm{k},\bm{u},\omega)
      =\frac{1}{i\bm{k}\cdot\bm{u}+\ell_e^\star(\omega)^{-1}}\left[\vphantom{\int}
         S(\omega)\right.
\\
         \left.+\frac{1}{\ell_s(\omega)}\int_{4\pi}p(\bm{u}\cdot\bm{u}',\omega)I(\bm{k},\bm{u}',\omega)\ud\bm{u}'
      \right].
\end{multline}
Transforming back to real space, and noting that $1/\alpha=\int_0^{+\infty}\exp(-\alpha s)\ud s$, we
obtain
\begin{multline}
   I(\bm{r},\bm{u},\omega)
      =\int_0^{+\infty}e^{-s/\ell_e^\star(\omega)}\left\{\vphantom{\int}
         S(\omega)\delta(s\bm{u})\right.
\\
         \left.+
         \frac{1}{\ell_s(\omega)}\int_{4\pi}p(\bm{u}\cdot\bm{u}',\omega)I(\bm{r}-s\bm{u},\bm{u}',\omega)\ud\bm{u}'
      \right\}\ud s
\end{multline}
which is the integral form of the RTE. By iterating this equation, we observe that light propagation
can be seen as a random walk process with a step-size distribution given by
$p(s,\omega)=\ell_s(\omega)^{-1}\exp[-s/\ell_s(\omega)]$, and an angular redistribution at each
scattering event given by $p(\bm{u}\cdot\bm{u}',\omega)$. Along a given path, photons can be absorbed and
emitted by the molecules, which is taken into account by the factor
$\exp[-\rho_0\sigma_a(\omega)s+\rho_1\sigma_{em}(\omega)s]$ in the integral.

With this picture in mind, the detected flux can be formally written as
\begin{equation}\label{integral}
   I(\omega)=\hbar\omega S(\omega)\int_{0}^{+\infty}P(s,\omega)
      e^{-\rho_0\sigma_a(\omega)s+\rho_1\sigma_{em}(\omega)s}\ud s
\end{equation}
where $P(s,\omega)$ is the path length distribution inside the scattering medium, that depends on
the geometry of the cell, the step-size distribution $p(s,\omega)$ and the phase function
$p(\bm{u}\cdot\bm{u}',\omega)$. We note that the origin of a path can be arbitrary since light is
emitted by spontaneous emission processes distributed in the volume of the scattering medium.

Following the main text, we now introduce the effective absorption length
\begin{equation}
   \frac{1}{\ell_a^\star(\omega)} =\frac{1}{\ell_e^\star(\omega)}-\frac{1}{\ell_s(\omega)}
\end{equation}
and consider that the average path length $\bra s(\omega)\ket=\int_0^{+\infty} sP(s,\omega)\ud s$ is
large compared to $\ell_a^\star$. In a strongly scattering medium, this implies that the width of
the distribution $P(s,\omega)$ is also large compared to $\ell_a^\star$, and we can assume
$P(s,\omega)$ to be uniform over a range of path lengths $s\in[0,s_m(\omega)]$ where
$s_m(\omega)\gtrsim\ell_a^\star$.  In that case, the integral in Eq.~\eqref{integral} can be easily
evaluated, and we find that
\begin{equation}
   I(\omega)=
   \frac{\eta\,\hbar\omega\mathcal{N}(\omega)}{\displaystyle\frac{\rho_0\sigma_a(\omega)}{\rho_1\sigma_{em}(\omega)}-1}.
\end{equation}
Finally, by using the Kennard-Stepanov relation and the definition of the chemical potential $\mu$ in the main text,
\begin{equation}
   I(\omega)=\frac{\eta\,\hbar\omega\mathcal{N}(\omega)}{\exp\displaystyle\left(\frac{\hbar\omega-\mu}{k_BT}\right)-1}
\end{equation}
which is Eq.~\eqref{eq:exp_intensity} of the main text.  In the diffusive regime, the average length $\bra
s(\omega)\ket$ can be estimated to be $3\alpha R^2/\ell_t$, and the thermalization condition $\bra
s(\omega)\ket\gg\ell_a^\star$ becomes identical to that in the main text.

\section{Methods}\label{app_methods}
% ===============

\subsection{Experimental setup}\label{app_setup}
% -----------------------------

A continuous-wave laser (Roithner Lasertechnik, $\lambda = \SI{473}{nm}$) is used to excite the
sample. The diameter of the beam is set to $\SI{8}{mm}$. The beam passes through a 50/50 polarizing
beam splitter and is divided into two orthogonally polarized components that excite the sample from
two opposite sides. The light emitted by the sample is collected, in a direction orthogonal to the
excitation beams, through a $4f$ optical system and is injected in the input fiber of a grating
spectrometer (ANDOR Kymera 328i) equipped with an EMCCD camera Andor iXon Ultra \& Life 888. For the
data presented in this manuscript the camera was used in the normal CCD regime, without
amplification gain.  The height of the collection system corresponds to the center of the excitation
beams.

For the measurements taken at temperatures different from ambient temperature, the sample is
installed on a flat heating resistance. A calibration curve converting voltage to temperature is
acquired, in the same conditions as the optical measurements presented in the main body of the
manuscript. A thermocouple placed inside the sample is used to measure the temperature achieved
after a time of 20 minutes over which the temperature of the sample stabilizes.  

\subsection{Sample preparation}\label{app_sample}
% -----------------------------

Samples are aqueous suspensions with variable concentrations of Rhodamine 6G and polystyrene
microbeads (Polysciences, Polybeads Microspheres $\SI{0.37}{\mu m}$). The polystyrene microbeads are
supplied as a $\SI{2.7}{\%}$ solids (w/v) aqueous suspension. First, the original bead solution is
placed in an ultrasonic bath for ten minutes to prevent the formation of aggregates. Then,  the
appropriate volume of the original bead solution diluted with milli-Q water to reach the desired
scatterers' concentration. 

Rhodamine 6G is supplied as a powder (Polysciences). To achieve the desired concentration, we
undergo a multi-step dilution process in milli-Q water. A solution of Rhodamine 6G at $\SI{5}{g/L}$
is placed in an ultrasonic bath for ten minutes to prevent molecular aggregation. Subsequently, the
appropriate volume can be taken and added to the diluted bead solution to obtain the desired final
dye concentration.

Before the experiment the sample is placed in an ultrasonic bath for fifteen minutes to prevent the
formation of aggregates. 

\subsection{Data analysis}\label{app_data}
% ------------------------

Spectra are obtained by averaging the vertical pixels of the spectrometer's camera as in standard
spectroscopy procedures. The spectrometer wavelength is calibrated with a mercury lamp. For each
spectrum measured on an active sample, we also acquire the spectrum of the background, on a sample
with the same concentration of scatterers as the sample of interest, without Rhodamine. The
background spectrum is acquired in the same experimental conditions (laser power, sample temperature
and acquisition settings for the camera and the spectrometer) as for the sample of interest. The
spectrum of the background is first rescaled such that its mean value in the spectral range 485-490
nm matches that of the active sample and then is subtracted to the spectrum of the active sample.
This procedure ensures a zero baseline in this spectral region in which fluorescent emission is
absent. The spectrometer’s wavelength resolution is \SI{0.8}{nm}, which corresponds to three
(horizontal) pixels of the camera; therefore, during post-processing, the intensities of three
consecutive pixels are averaged. Then, each spectrum is normalized by the total acquisition time.
Since the definition of $\Theta$ involves a numerical derivative operation on the logarithm of the
spectrum, that is a procedure that adds numerical noise to the data, the logarithm of each spectrum
is first smoothed with a moving average over a \SI{3}{nm} window before differentiation. 

% Bibliography

\end{document}